\documentclass[twocolumn,10pt]{jmlr} 

\usepackage{booktabs}
\usepackage{subcaption}
\usepackage{siunitx}

\usepackage[switch]{lineno}

\theorembodyfont{\upshape}
\theoremheaderfont{\scshape}
\theorempostheader{:}
\theoremsep{\newline}

\title[Biomedical Retrieval Pipeline Trade-offs]{A Systematic Study of Biomedical Retrieval Pipeline Trade-offs in Performance and Efficiency}

\author{%
\Name{Hayk Stepanyan} \Email{hs3627@columbia.edu}\\
\addr Columbia University, United States of America
\AND
\Name{Matthew McDermott} \Email{mm6677@cumc.columbia.edu}\\
\addr Columbia University, United States of America
}

\makeatletter

\def\ps@jmlrtps{%
  \def\@oddhead{}%
  \def\@evenhead{}%
  \def\@oddfoot{\hfil\thepage\hfil}%
  \def\@evenfoot{\hfil\thepage\hfil}}

\makeatother
\begin{document}

\maketitle


\begin{abstract}
Retrieval systems are increasingly used in biomedical and clinical natural language processing applications, yet practical guidance for researchers building such systems is limited. In this work, we provide such guidance through an empirical study of how retrieval pipeline design choices affect performance and efficiency at scale.

In particular, we examine retrieval over a variety of existing, public biomedical text datasets, leveraging a variety of disparate types of queries, including exam-style questions, conversational medical queries, community-asked questions, and non-question formulations across various retrieval pipeline settings spanning corpus selection, chunk granularity, and vector index configuration. Retrieval results are judged using a robust, win-rate comparison assessment via an LLM-as-a-judge setting with human validation.

Across these experiments, we identify several points of concrete guidance for reviewers, including the superiority of corpus aggregation for absolute retrieval quality, and the emergence of MedRAG/pubmed as the Pareto-optimal singleton corpus under graph-based (HNSW) indexing, appropriate chunking strategies, and FAISS indexing choices that offer the best trade-offs in speed and efficiency.
\end{abstract}

\paragraph*{Data and Code Availability}
This work uses publicly available biomedical text and question datasets hosted on Hugging Face. Specifically, we use the MedRAG corpora (MedRAG/textbooks, MedRAG/pubmed, and MedRAG/wikipedia), the PubMed Central Open Access subset (pmc/open access), and multiple biomedical question and conversational datasets (MedMCQA, MedQuad, BioLeaflets, ChatDoctor). All datasets are publicly accessible under their respective licenses and are available to other researchers via the Hugging Face repository.

The code used to construct the retrieval pipelines is available at \url{https://github.com/McDermottHealthAI/Medical-Retrieval-DB}.

\paragraph*{Institutional Review Board (IRB)} This work does not require an IRB approval.

\section{Introduction}
\label{sec:intro}
High-quality retrieval is increasingly critical in clinical and biomedical natural language processing (NLP) applications, where clinical decision-making and medical research rely on grounding predictions in a vast and continually growing literature~\citep{Krithara2023BioASQ}. Retrieval-augmented systems have been shown to improve factual accuracy and interpretability~\citep{Lewis2020RAG,izacard-grave-2021-leveraging}, but their performance is highly sensitive to upstream retrieval design decisions~\citep{Thakur2021BEIR}.

\begin{figure}[t]
\centering
\includegraphics[width=\linewidth]{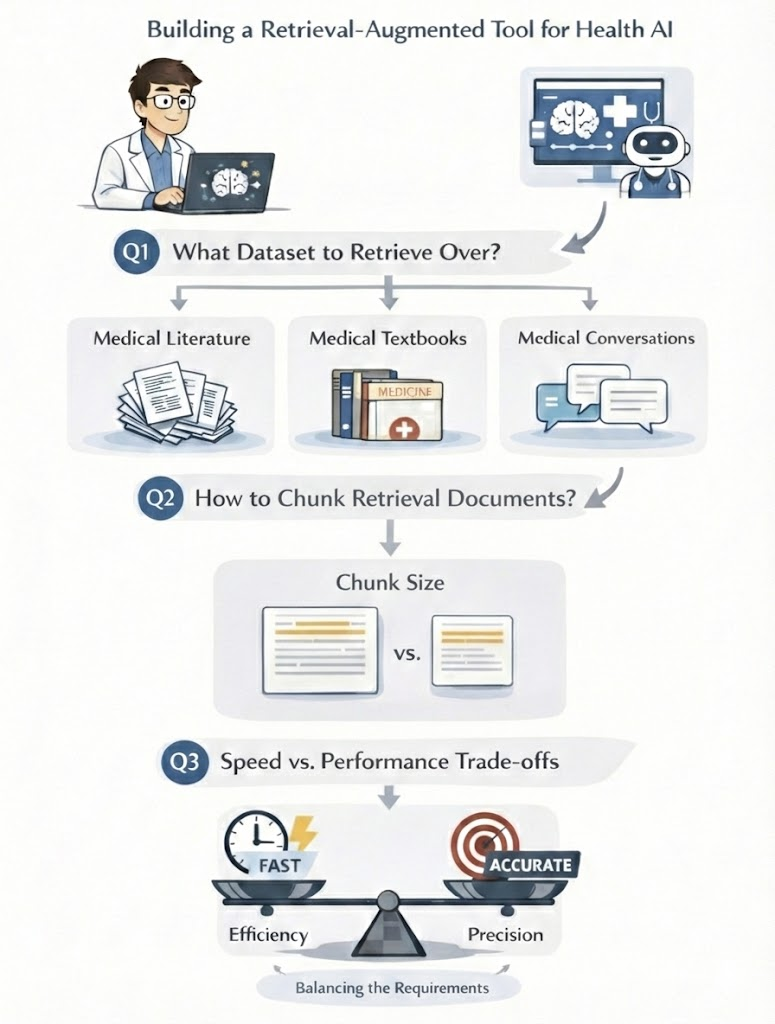}
\caption{
Overview of the retrieval design space and the three fundamental research questions studied in this work.
}
\label{fig:design_overview}
\end{figure}

Despite the availability of general-purpose retrieval tools and frameworks~\cite{gao2024retrievalaugmentedgenerationlargelanguage}, most biomedical retrieval pipelines are assembled in an ad hoc manner. \textit{As a result, practitioners lack clear guidance, often leading to suboptimal design choices in real-world systems} \citep{liu2025improving}. In particular, we identify the following three fundamental questions, illustrated in Figure~\ref{fig:design_overview}, that researchers leveraging retrieval must answer when designing their systems:

\begin{enumerate}
    \item \textbf{Corpus selection:} Which publicly available biomedical datasets provide the strongest retrieval performance across diverse medical query types?
    \item \textbf{Chunking granularity:} How does document segmentation affect retrieval quality?
    \item \textbf{Efficiency-performance trade-offs:} How do retrieval system configurations—including indexing, chunking, and corpus scale—affect the trade-off between retrieval quality and computational efficiency?
\end{enumerate}
In this work, we address these questions empirically across a wide range of retrieval settings, with the goal of providing practical and actionable guidance for researchers and practitioners.
The remainder of the paper reviews related work, defines the scope of our retrieval analysis, presents empirical results corresponding to each design question, and concludes with a discussion of limitations and future work. We find that corpus selection dominates retrieval performance: aggregating biomedical sources consistently yields the highest quality, while under graph-based FAISS indexing (HNSW), MedRAG/pubmed provides the strongest singleton Pareto trade-off between quality and throughput We also find that chunking granularity has no universally optimal setting but instead depends strongly on query type, and that HNSW-based FAISS indices offer the most favorable quality–efficiency trade-offs for dense biomedical retrieval.

\section{Related Work}
Biomedical question answering systems commonly rely on retrieval over large-scale corpora such as PubMed, PubMed Central, Wikipedia, and curated medical resources. Benchmarks including BioASQ \citep{Krithara2023BioASQ}, MedQA \citep{jin2020diseasedoespatienthave}, and MedMCQA \citep{pmlr-v174-pal22a} have driven progress in this area by evaluating end-to-end QA performance, while recent retrieval-augmented generation (RAG) systems integrate dense retrievers with language models to ground outputs in biomedical literature \citep{li2025biomedrag, xiong-etal-2024-benchmarking}.

A growing body of work proposes specialized biomedical RAG frameworks, including MedRAG~\citep{medrag}, BiomedRAG~\citep{li2025biomedrag}, Self-BioRAG~\citep{selfbiorag}, and i-MedRAG~\citep{Xiong2025}, which introduce tailored retrieval mechanisms such as iterative querying, re-ranking, and self-reflection to improve factual grounding and reduce hallucinations. Surveys further categorize these approaches and document their rapid adoption in medical and clinical settings~\citep{he2025retrievalaugmentedgenerationbiomedicinesurvey, Neha2025}. However, in most cases, the retriever is treated as a fixed component, and retrieval design choices are evaluated only through downstream QA or generation metrics.

Several recent studies have begun to analyze retrieval-related trade-offs in RAG systems. Prior work examines latency–accuracy trade-offs in large-scale RAG inference~\citep{shen2024towards}, evaluates alternative retrieval strategies for medical chatbots~\citep{bora2024systematic}, and explores hybrid lexical–semantic retrieval pipelines on biomedical corpora~\citep{Stuhlmann_2025}. Related efforts also investigate re-ranking architectures, such as ModernBERT + ColBERT ~\citep{rivera2025modernbertcolbertenhancing}, to improve retrieval quality while maintaining efficiency. While these studies highlight the importance of retrieval efficiency and design, they do not systematically characterize how core retrieval design choices—such as corpus selection, chunking granularity, and indexing configuration—interact to shape retrieval quality and efficiency in biomedical settings, which is the focus of this work.

\section{Problem Statement}
We study the problem of designing a high-quality dense retrieval component for biomedical text. Within a fixed single-stage dense retrieval pipeline, practitioners must make several interacting design decisions—corpus selection, document chunking, and indexing strategies—that materially affect retrieval quality.

Our focus is on precision-oriented retrieval: given a biomedical query, the system should return a small set of passages that are maximally useful for a user or a downstream reasoning model. 
To isolate the effects of core design choices, we restrict attention to single-stage dense retrieval and intentionally exclude multi-stage pipelines, re-ranking, hybrid sparse–dense methods, and modifications to retrieval algorithms.

Rather than introducing new retrieval models, we empirically study how these existing design decisions shape retrieval quality and efficiency. 
Our goal is to characterize the trade-offs inherent in dense retrieval design and to provide practical, evidence-based guidance for constructing effective biomedical retrieval systems.

\section{Experimental Setup}
\label{sec:setup}

We study a single-stage dense retrieval setting over large biomedical text corpora. Given a natural language query $q$ and a corpus $\mathcal{C}$ of biomedical documents, the retriever returns the top-$k$ text chunks ranked by vector similarity between query and document embeddings.

This setup reflects a common retrieval component in biomedical question answering and decision-support systems \citep{Yang2025}, where a small number of relevant passages must be selected from millions of candidates to support downstream reasoning. We focus on single-stage dense retrieval to isolate the effects of representation, chunking, indexing, and corpus design, deliberately abstracting away multi-stage, hybrid, and re-ranking architectures whose additional components confound these factors.

\subsection{Query Distributions}
\label{subsec:queries}

Retrieval behavior depends strongly on query formulation. We therefore evaluate retrieval using a balanced set of queries drawn from ten distinct sources, summarized in Table~\ref{tab:queries}. These sources span exam-style medical questions (MedMCQA \citep{pmlr-v174-pal22a}, MedQA \citep{jin2020diseasedoespatienthave}), conversational queries (ChatDoctor \citep{li2023chatdoctormedicalchatmodel}), community-asked health questions (Health StackExchange \citep{healthStackExchange}), scientific queries derived from biomedical literature (PubMed titles \citep{xiong2024benchmarking} and PMC question sentences \citep{PMCOpenAccess}), keyword-style queries (MeSH terms \citep{MeSH2025} and Wikipedia medical terms \citep{xiong2024benchmarking}), and novel LLM-generated queries\footnote{All generated queries will be made publicly available upon publication.} in both question–answer and non-question formats. Example queries are provided in Appendix \ref{apd:queries}.

\subsection{Biomedical Corpora}
\label{subsec:corpora}

We evaluate retrieval over several publicly available biomedical text corpora that vary in scale, curation, and intended use. These include curated educational resources (MedRAG/textbooks), large-scale biomedical research collections (MedRAG/pubmed and PMC Open Access), and general-purpose medical reference sources (MedRAG/wikipedia), as well as smaller, domain-specific datasets commonly used in prior biomedical QA work (ChatDoctor, MedMCQA, BioLeaflets \citep{yermakov-etal-2021-biomedical}, MedQuad \citep{MedQuAD_HuggingFace}).

Table~\ref{tab:datasets} summarizes the datasets used in this study, including the number of documents, total token count, and approximate storage footprint after chunking.

\begin{table*}[t]
\centering
\floatconts
  {tab:queries}
  {\caption{Query sources used for retriever evaluation, grouped by high-level query category. 25 queries per source are used to make up our total query set.}}
  {\begin{tabular}{lll}
  \toprule
  \bfseries Category & \bfseries Query Source & \bfseries Description \\
  \midrule
  Exam-style &
  MedMCQA &
  Indian postgraduate medical exam questions \\
  &
  MedQA &
  Standardized medical licensing questions \\
  \addlinespace

  Conversational &
  ChatDoctor &
  Doctor--patient conversational queries \\
  &
  Health StackExchange &
  Community-driven health questions \\
  &
  LLM-generated (Q/A) &
  Gemini-generated medical Q/A-style queries \\
  \addlinespace

  Scientific &
  PubMed Titles &
  Biomedical research article titles \\
  &
  PMC Question Sentences &
  Question sentences extracted from PMC articles \\
  &
  LLM-generated (Non-QA) &
  ChatGPT-generated non-question queries \\
  \addlinespace

  Keyword-based &
  MeSH Terms &
  Controlled medical subject headings \\
  &
  Wikipedia Medical Terms &
  Medical concepts from Wikipedia \\
  \bottomrule
  \end{tabular}}
\end{table*}

\begin{table*}[t]
\centering
\floatconts
  {tab:datasets}
  {\caption{Biomedical datasets used in this work. We report the number of documents, total token count, and approximate storage size for each dataset. Reported statistics are based on 512-token text chunks embedded using the Qwen/Qwen3-Embedding-0.6B model.}}
  {\begin{tabular}{lrrr}
  \toprule
  \bfseries Dataset & \bfseries \# Documents & \bfseries \# Tokens & \bfseries Size (GB) \\
  \midrule
  MedRAG/wikipedia & 29{,}803{,}333 & 4{,}934{,}754{,}781 & 126 \\
  MedRAG/pubmed & 17{,}067{,}629 & 4{,}841{,}492{,}054 & 78 \\
  PMC Open Access & 11{,}360{,}871 & 5{,}677{,}602{,}919 & 55 \\
  MedMCQA & 187{,}872 & 23{,}935{,}963 & 0.81 \\
  ChatDoctor & 113{,}196 & 25{,}798{,}748 & 0.51 \\
  MedRAG/textbooks & 125{,}914 & 22{,}899{,}703 & 0.54 \\
  BioLeaflets & 69{,}088 & 35{,}173{,}307 & 0.33 \\
  MedQuad & 19{,}221 & 4{,}700{,}614 & 0.09 \\
  \bottomrule
  \end{tabular}}
\end{table*}

\subsection{Experimental Retrieval Pipeline}
\label{subsec:pipeline}
We use a modular dense retrieval pipeline to enable controlled comparisons across retrieval design choices. The pipeline decomposes retrieval into three components: (i) chunking and embedding, (ii) indexing, and (iii) nearest-neighbor retrieval. While the pipeline itself is not the object of study, it provides a consistent experimental framework in which individual components can be varied independently while others are held fixed.

\paragraph{Chunking and Embedding.}
\label{subsubsec:chunking}

Biomedical documents are segmented into fixed-length text chunks using a configurable token budget. We vary the chunk size to study how retrievable unit granularity affects retrieval quality across different query types.

Each chunk is encoded into a dense vector using transformer-based embedding models accessed via the SentenceTransformers~\citep{reimers2019sentencebert} library. We evaluate both general-purpose and biomedical-specific embedding models. Queries are embedded using the same model as the corpus to ensure representational consistency.

\paragraph{Indexing.}
\label{subsubsec:indexing}

Dense vectors are indexed using FAISS~\citep{Johnson2019FAISS}, as implemented in the Hugging Face Datasets library \citep{lhoest-etal-2021-datasets}, to support scalable nearest-neighbor search. We evaluate multiple FAISS index configurations, including exact and approximate indices, to characterize trade-offs between retrieval quality and latency.

Index configurations are specified using FAISS string factories, enabling consistent index construction across experimental runs. By fixing the corpus, chunking strategy, and embedding model while varying the index type, we analyze the trade-offs between retrieval effectiveness and computational efficiency.

\paragraph{Retrieval.}
\label{subsubsec:retrieval}

At query time, each query is embedded using the same model applied during corpus construction. Queries are not chunked, even if they exceed the retrieval corpus chunk size. Retrieval is performed by querying the FAISS index to obtain the top-$k$ nearest-neighbor text chunks under vector similarity.

We retrieve the top-$5$ results per query, reflecting a common downstream setting in which a small number of retrieved passages is provided to a reasoning or generation model.

Unless otherwise specified, all experiments use single-stage dense retrieval with 512-token chunks, the Qwen/Qwen3-Embedding-0.6B embedding \citep{qwen3embedding} model, and an exact FAISS \texttt{IndexFlat} configuration.

\subsection{Evaluation Methods}
Our evaluation focuses on a user-centric question: \emph{given two retrieval configurations, which system returns passages that a user would prefer for a given biomedical query?} In this way, rather than evaluating absolute relevance scores or precision-based metrics, we compare retrieval systems directly based on the relative usefulness of their retrieved results.

Given two retrieval settings (e.g., differing in corpus, chunking, or indexing), we compare their outputs for the same query using rank-aligned, passage-level comparisons. Specifically, for each rank $i \in \{1, \dots, k\}$, the passage retrieved at rank $i$ under one setting is compared against the passage retrieved at the same rank under the other setting.

An LLM is prompted to judge which of the two passages is more helpful for answering the query, or to indicate a tie if neither passage is clearly preferable. This procedure directly operationalizes user preference between competing retrieval configurations.

Our primary evaluation metric is the \emph{win rate}. For a pair of retrieval settings $A$ and $B$, the win rate of $A$ over $B$ is defined as the fraction of passage-level comparisons in which $A$ is preferred to $B$. Each query contributes up to $k$ independent comparisons, corresponding to the top-$k$ retrieved passages.

We choose this metric because it directly approximates the fraction of times a user would prefer the retrieved output of system $A$ over system $B$ for a given query \citep{win_rate}.

We validate the LLM-based judgments by measuring agreement with expert human annotations on a random subset of comparisons, consistently finding high agreement (see Appendix~\ref{apd:human_validation} for more details).

\section{Results}
In this section, we present results for the three key design decisions of the retrieval pipeline: (1) which biomedical corpora are most effective for large-scale retrieval, 
(2) how chunking granularity affects retrieval quality, and 
(3) what are the trade-offs between retrieval performance and efficiency. 
\subsection{Corpus Selection for Biomedical Retrieval}
\label{subsec:corpus-analysis}

\begin{figure*}[t]
\centering
\includegraphics[width=0.95\textwidth]{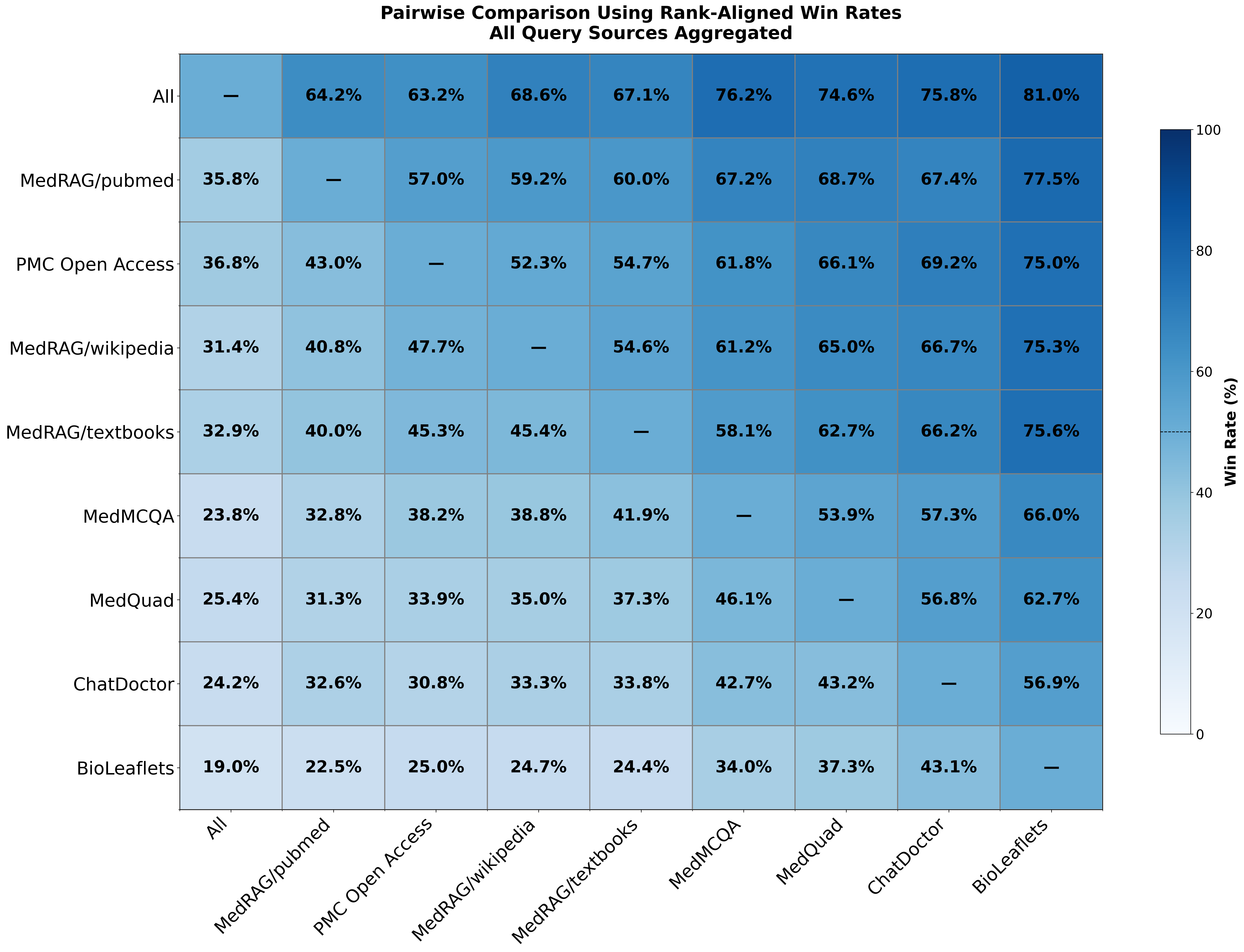}
\caption{Pairwise corpus comparison using rank-aligned win rates aggregated across all query sources. Each cell shows the proportion when the row corpus outperforms the column corpus. Draw rates are reported in Appendix \ref{apd:first}.}
\label{fig:corpus_tournament}
\end{figure*}

\emph{Corpus choice is the dominant factor governing retrieval quality, outweighing chunking and indexing decisions. Across all query types, the aggregated corpus consistently outperforms every individual dataset, achieving a win rate above 63\% against all single-corpus configurations (Figure \ref{fig:corpus_tournament})}. Aggregation improves retrieval robustness by increasing coverage across heterogeneous query formulations, from exam-style questions to conversational and synthetic queries. This result is expected: in principle, a sufficiently expressive embedding model operating over the union of all datasets should be able to retrieve relevant content better than a singleton dataset.

\begin{figure*}[t]
\centering
\includegraphics[width=0.95\textwidth]{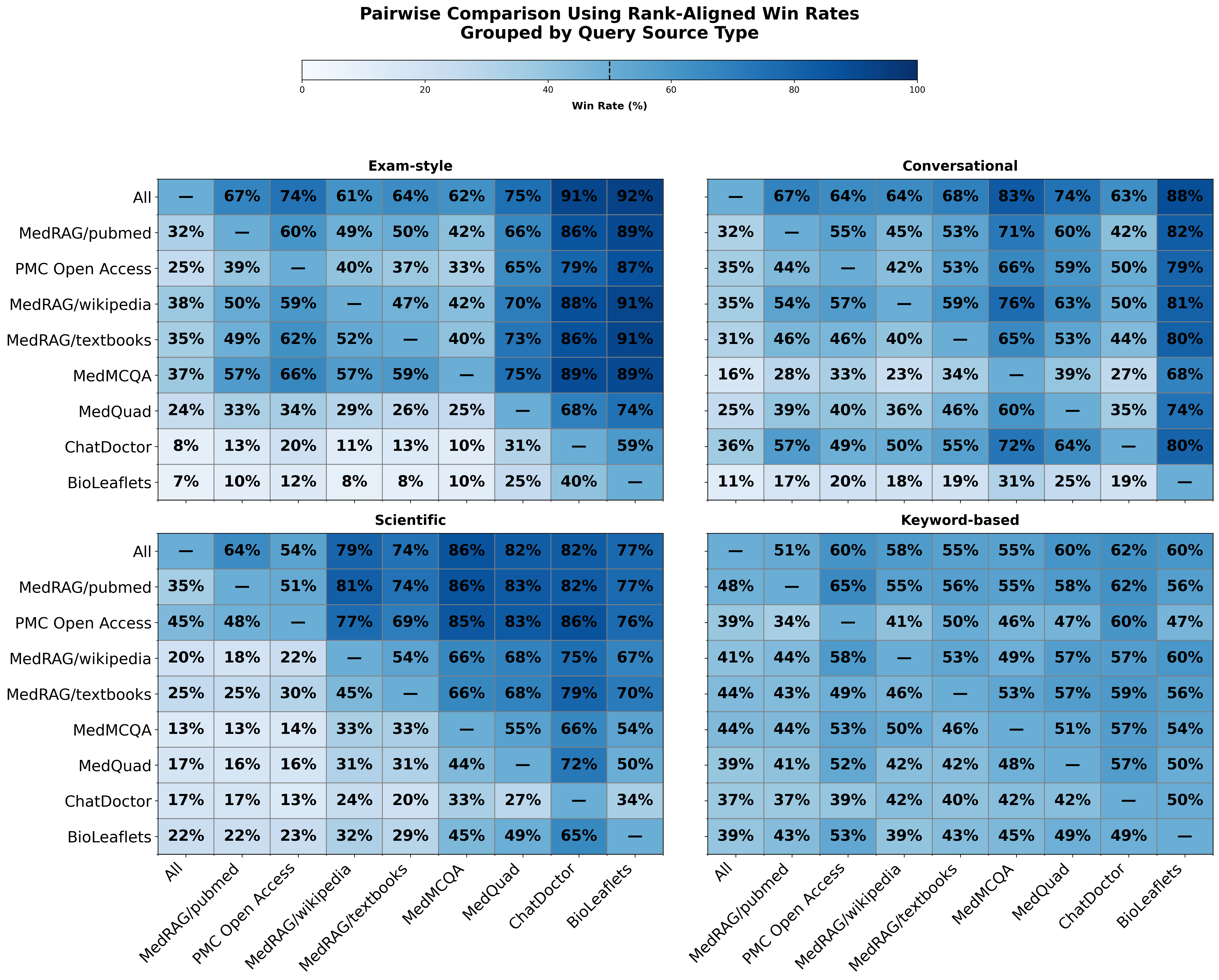}
\caption{Pairwise corpus comparison across query categories showing that while the aggregated "All" corpus provides the highest absolute quality, MedRAG/textbooks dominate exam-style and Wikipedia is preferred for specialized conversational queries.}
\label{fig:corpus_tournament_by_groups}
\end{figure*}

Retrieval performance is also strongly query-dependent. MedRAG/textbooks dominate technical and exam-style queries, and Wikipedia is preferred for specialized conversational queries, as shown in Figure \ref{fig:corpus_tournament_by_groups}.

\subsection{Impact of Chunking Granularity on Retrieval Quality}
\label{subsec:chunk-analysis}
Chunking granularity has no universally optimal setting; its effect on retrieval quality is strongly conditioned on query intent.

When all queries are aggregated together, retrieval quality is largely insensitive to chunk size. Aggregate win rates remain close to chance across granularities ranging from 64 to 2048 tokens (48.4\%–51.6\%), with high tie rates indicating substantial semantic overlap between adjacent chunk sizes (Figure \ref{fig:granularities_tournament}). This suggests that, on average, different segmentations often surface similar evidence.

\begin{figure}[t]
\centering
\includegraphics[width=1.0\linewidth]{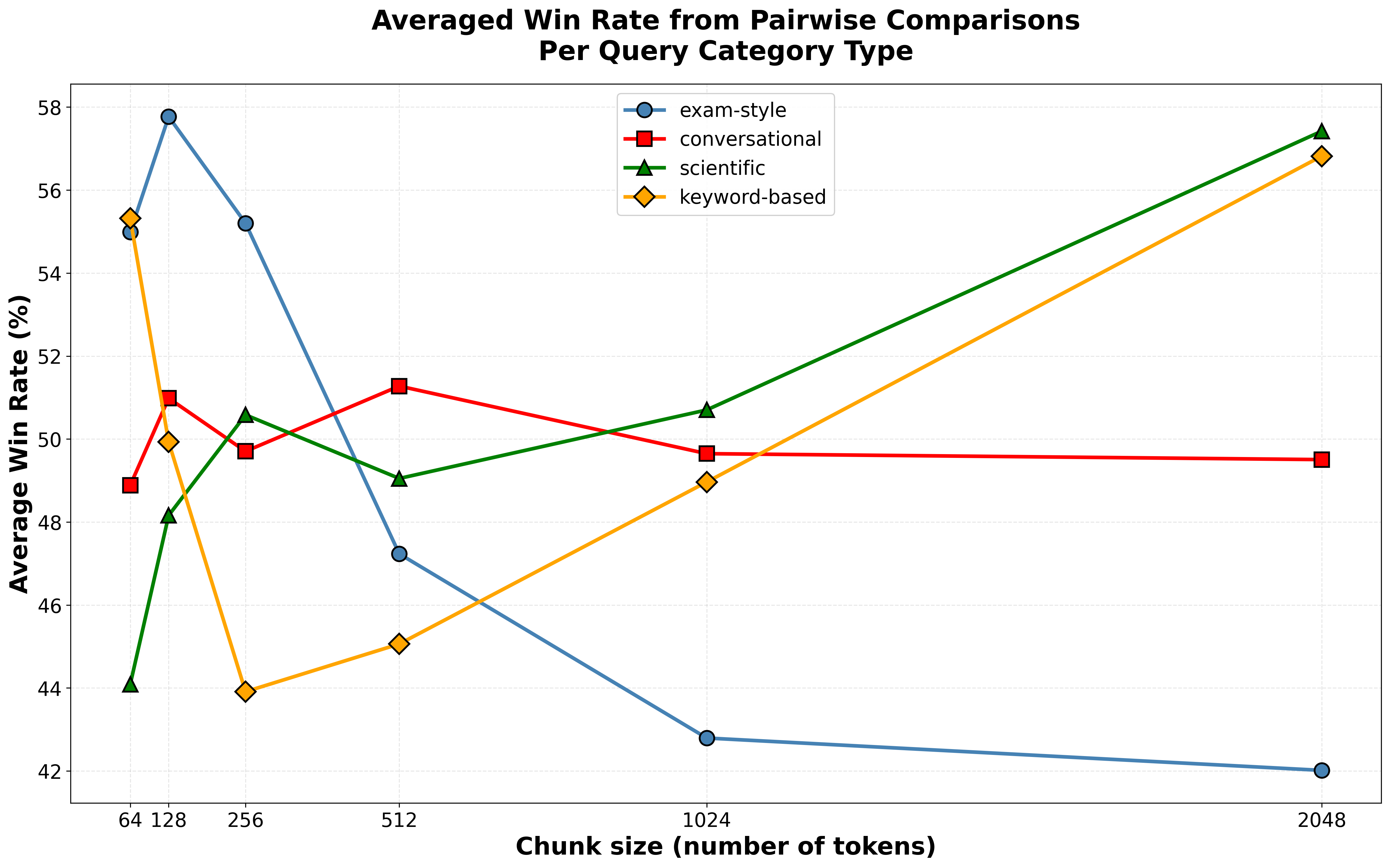}
\caption{
Averaged win rates across varying chunk sizes stratified by query category. The experiment is run on the PMC Open Access dataset only.
}
\label{fig:granularity_trends}
\end{figure}

However, query-specific trends emerge when results are stratified by query type, as shown in Figure \ref{fig:granularity_trends}. Exam-style and fact-heavy queries strongly favor small chunks, with performance peaking at 128 tokens and degrading monotonically as chunk size increases, consistent with the need for precise, undiluted factual evidence. In contrast, scientific queries benefit from larger chunks, with win rates increasing steadily up to 2048 tokens, indicating that broader contextual grounding improves retrieval for complex, research-oriented questions. Conversational and keyword-based queries exhibit weaker and less consistent preferences, with relatively flat or non-monotonic trends across granularities.

Retrieval systems should select chunk size based on expected query type—favoring smaller chunks for fact-heavy queries and larger chunks for scientific or context-rich queries—rather than adopting a single global default.

\subsection{Pareto Trade-offs Between Retrieval Quality and Efficiency}
Retrieval design choices induce clear quality–efficiency Pareto frontiers, revealing which configurations are strictly dominated under realistic throughput constraints. Overall, while corpus aggregation yields the highest absolute retrieval quality, MedRAG/textbooks is the optimal singleton corpus for high-quality biomedical retrieval under tight computational constraints.

We analyze the Pareto frontier between retrieval quality and system efficiency, highlighting the trade-offs imposed by corpus scale, chunking granularity, and vector index design.

\begin{figure}[t]
\centering
\includegraphics[width=1.0\linewidth]{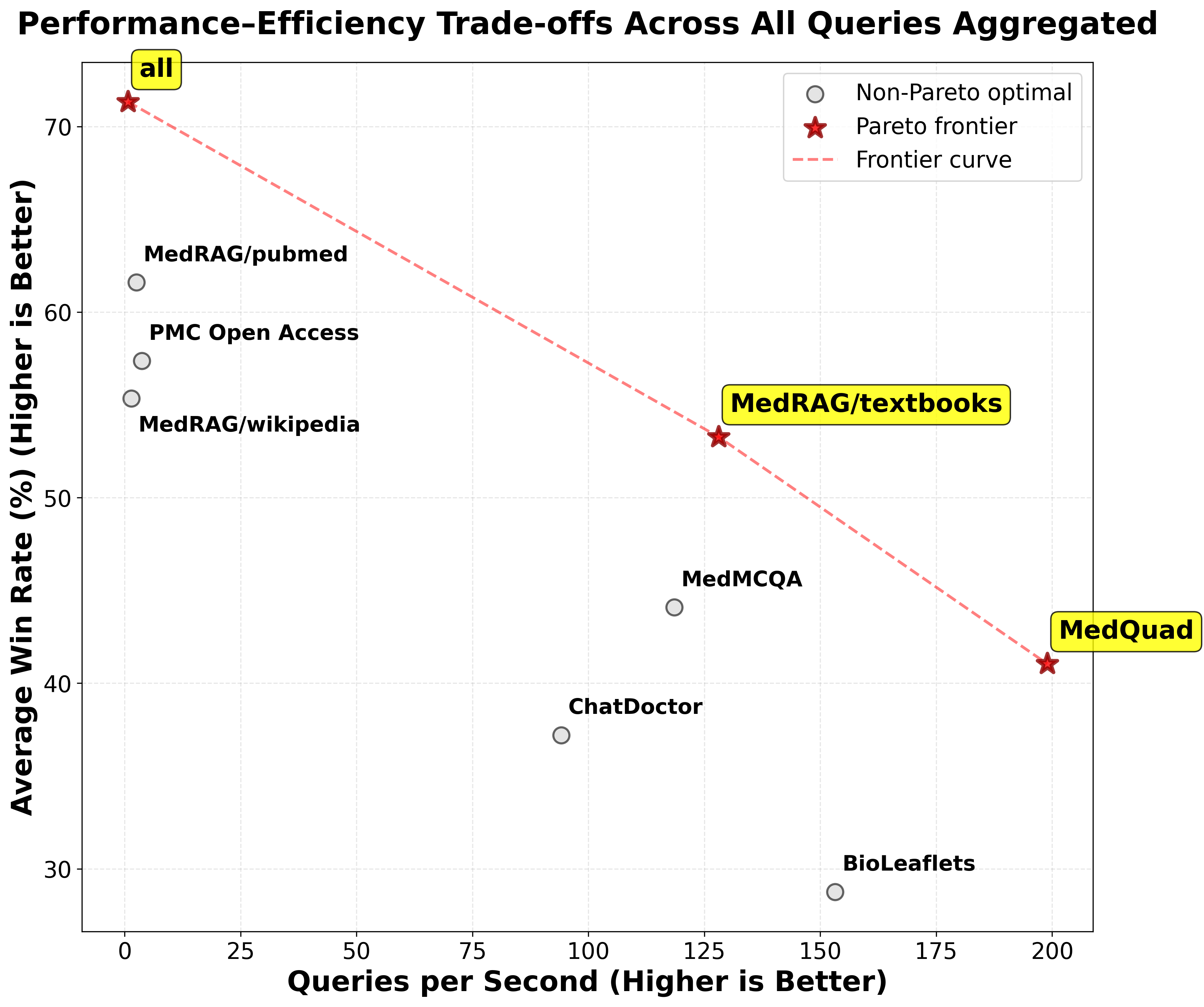}
\caption{
Pareto frontier illustrating the trade-off between average retrieval win rate and throughput (queries per second). This experiment is run using the IndexFLAT FAISS index.}
\label{fig:pareto_corpus}
\end{figure}

\paragraph{Corpus scale.} Under exact search (IndexFlat), the full All corpus achieves the highest retrieval quality ($\sim$70\% win rate) but operates at sub-1 QPS due to exhaustive nearest-neighbor search, placing it at the extreme quality end of the frontier (Figure~\ref{fig:pareto_corpus}). However, under HNSW64 indexing, throughput increases substantially and becomes far less sensitive to corpus size: the All corpus achieves approximately 50 QPS while retaining its quality advantage.

\begin{figure}[t]
\centering
\includegraphics[width=1.0\linewidth]{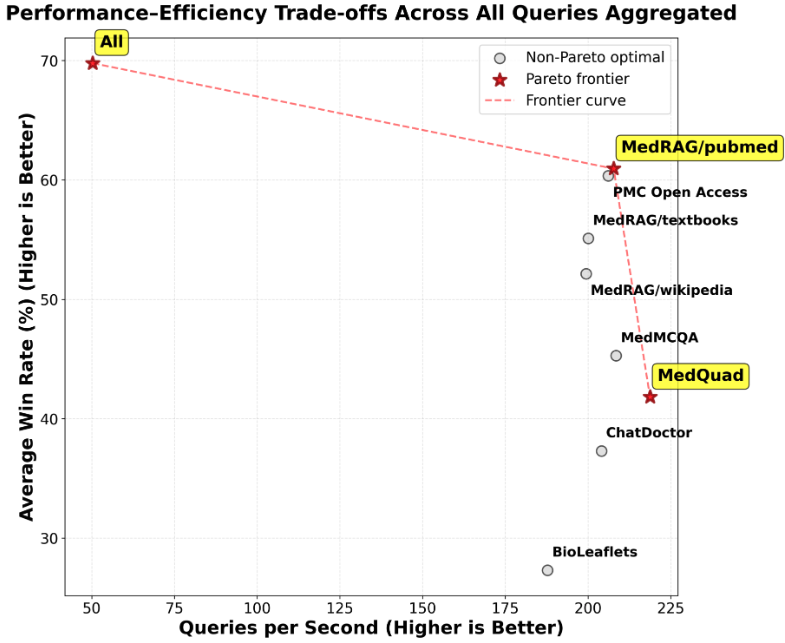}
\caption{
Pareto frontier illustrating the trade-off between average retrieval win rate and throughput (queries per second). This experiment is run using the HNSW64 FAISS index.}
\label{fig:pareto_corpus_hnsw64}
\end{figure}

This shift alters the Pareto landscape. Because graph-based search scales sublinearly with corpus size, larger corpora are no longer severely penalized in latency. As a result, MedRAG/pubmed moves onto the Pareto frontier, achieving higher retrieval quality than MedRAG/textbooks with only a modest throughput difference (Figure~\ref{fig:pareto_corpus_hnsw64}).

\paragraph{Chunking granularity.} Larger chunks reduce index size and search overhead. Across all queries, 2048-token chunks lie on the efficiency frontier, while smaller chunks (e.g., 128 tokens) achieve similar quality but incur an order-of-magnitude throughput penalty.

\paragraph{Vector Indexing.} IndexFLAT serves as the accuracy anchor, achieving the highest win rate ($\approx$ 59\%) at negligible throughput due to exhaustive search. Graph-based indices dominate the frontier: HNSW64 offers the best overall trade-off ($\approx$53.8\% win rate at $\sim$50 QPS), while HNSW16 maximizes throughput ($>$ 60QPS). The observed Pareto frontier closely matches expectations from the FAISS library paper. IVF and LSH-based methods fall below the frontier, consistent with FAISS guidance that quantization and inverted-file methods favor efficiency at very large scales but underperform graph-based methods on small-to-medium datasets.

Graph-based FAISS indices (e.g., HNSW variants) dominate the quality–efficiency frontier for dense biomedical retrieval at small-to-medium scales, while corpus and chunking choices primarily determine operating position along that frontier

\section{Limitation and Future Work}
\paragraph{Limitations.}
Our evaluation relies on LLM-based pairwise judgments, which, while validated against expert annotations, may not fully capture nuanced clinical preferences or downstream task performance. We focus exclusively on dense retrieval pipelines and do not evaluate hybrid or re-ranking-based systems. Additionally, our study centers on English-language biomedical corpora and does not address multilingual or low-resource settings.

\paragraph{Future Work.}
Future directions include adaptive retrieval systems that dynamically select corpora and chunking strategies based on inferred query intent, extending evaluation to hybrid sparse–dense retrieval and learned re-rankers, and integrating end-to-end downstream task performance such as clinical question answering or decision support. Expanding human validation and exploring multilingual biomedical retrieval remain important areas for further study.

\section{Conclusion}
We presented a systematic, user-centric evaluation of design choices in large-scale biomedical retrieval, focusing on corpus selection, chunking granularity, and indexing configurations. By framing evaluation as a direct preference comparison between retrieval systems, our study moves beyond traditional precision-based metrics and provides actionable guidance grounded in approximate user-perceived utility.

Our results show that corpus choice dominates retrieval performance. Aggregating multiple biomedical sources yields the strongest overall retrieval quality, while MedRAG/pubmed emerges as the most effective single-source corpus, offering high retrieval quality with substantially lower computational and storage cost. Chunking granularity has a weaker aggregate effect but exhibits strong query-dependent behavior: smaller chunks favor fact-heavy exam-style queries, whereas larger chunks improve retrieval for scientific and context-rich queries.

\begin{figure}[t]
\centering
\includegraphics[width=1.0\linewidth]{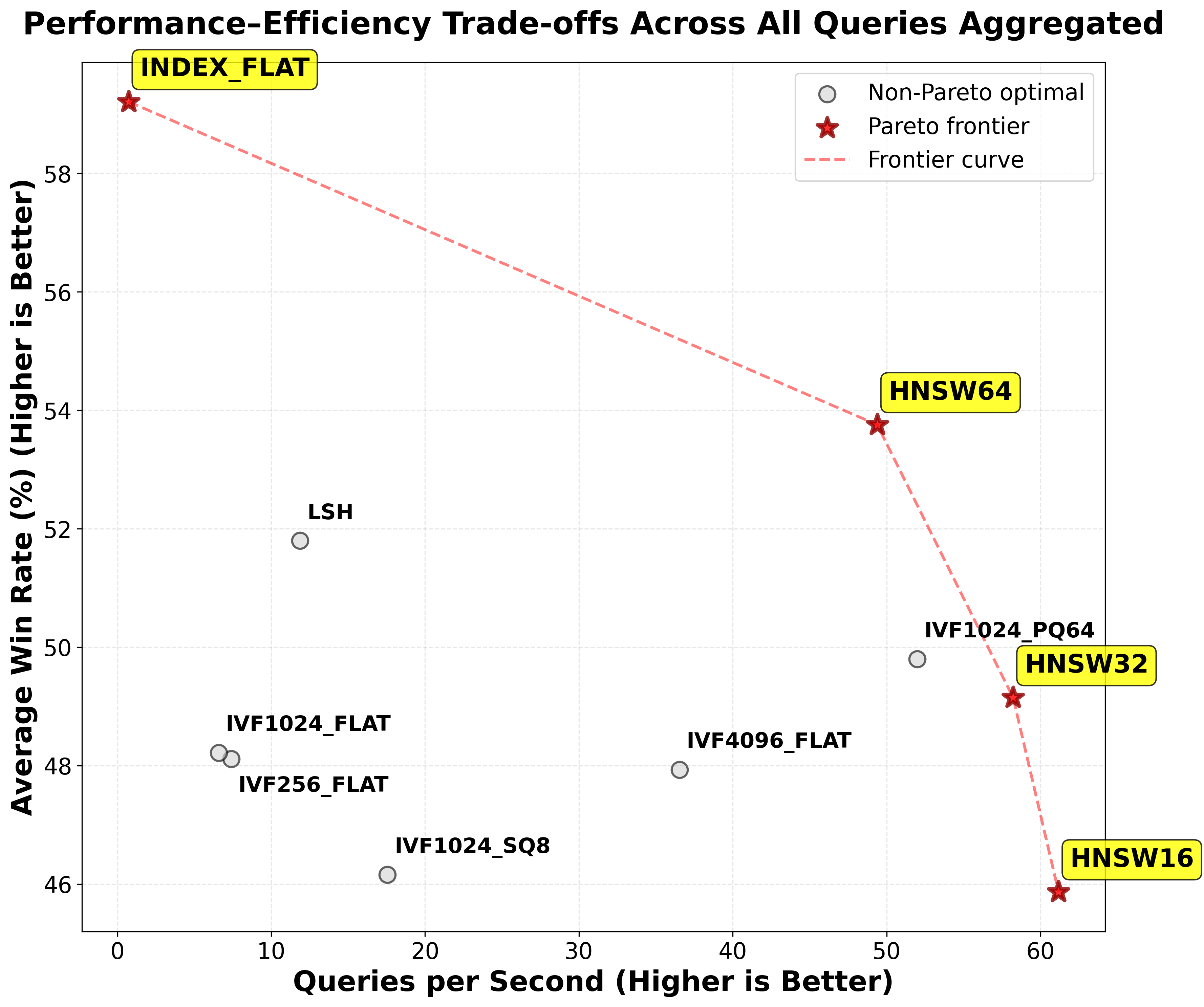}
\caption{
Pareto frontier illustrating the trade-off between average retrieval win rate and throughput (queries per second) across different FAISS indexes. The experiment is run on the aggregated All corpus.}
\label{fig:pareto_faiss}
\end{figure}

\bibliography{chil-sample}

\appendix

\section{Corpus Comparisons Plots}\label{apd:first}
\begin{figure}[t]
\centering
\includegraphics[width=1.0\linewidth]{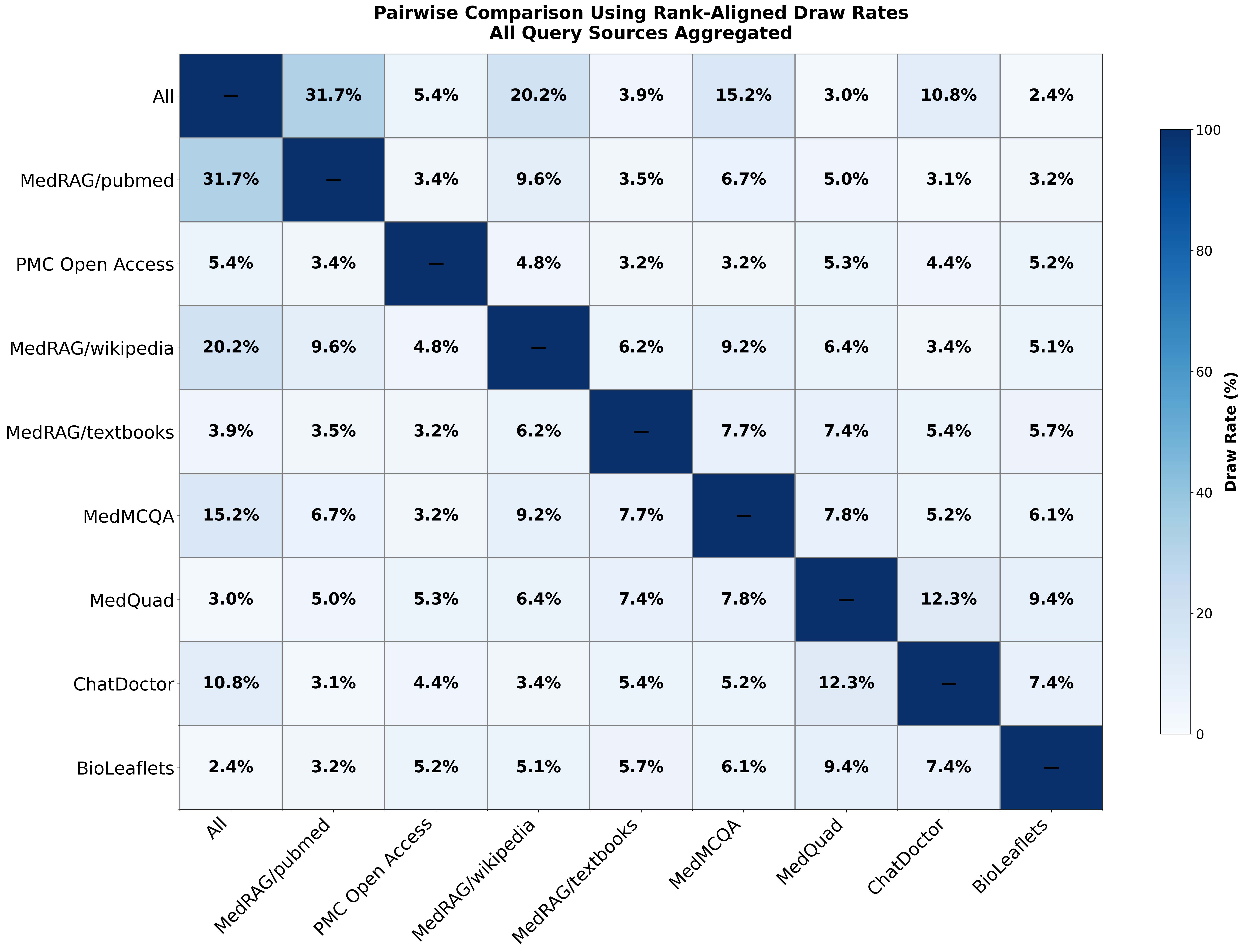}
\caption{
Pairwise draw rates for corpus comparison using rank-aligned win rates for all query sources.
}
\label{fig:corpus_draw}
\end{figure}

Figure \ref{fig:corpus_draw} represents the pairwise draw rates for the corpus-level retrieval comparisons.

\section{Chunk Comparisons Plots}
\label{apd:second}
\begin{figure}[t]
\centering
\includegraphics[width=0.95\linewidth]{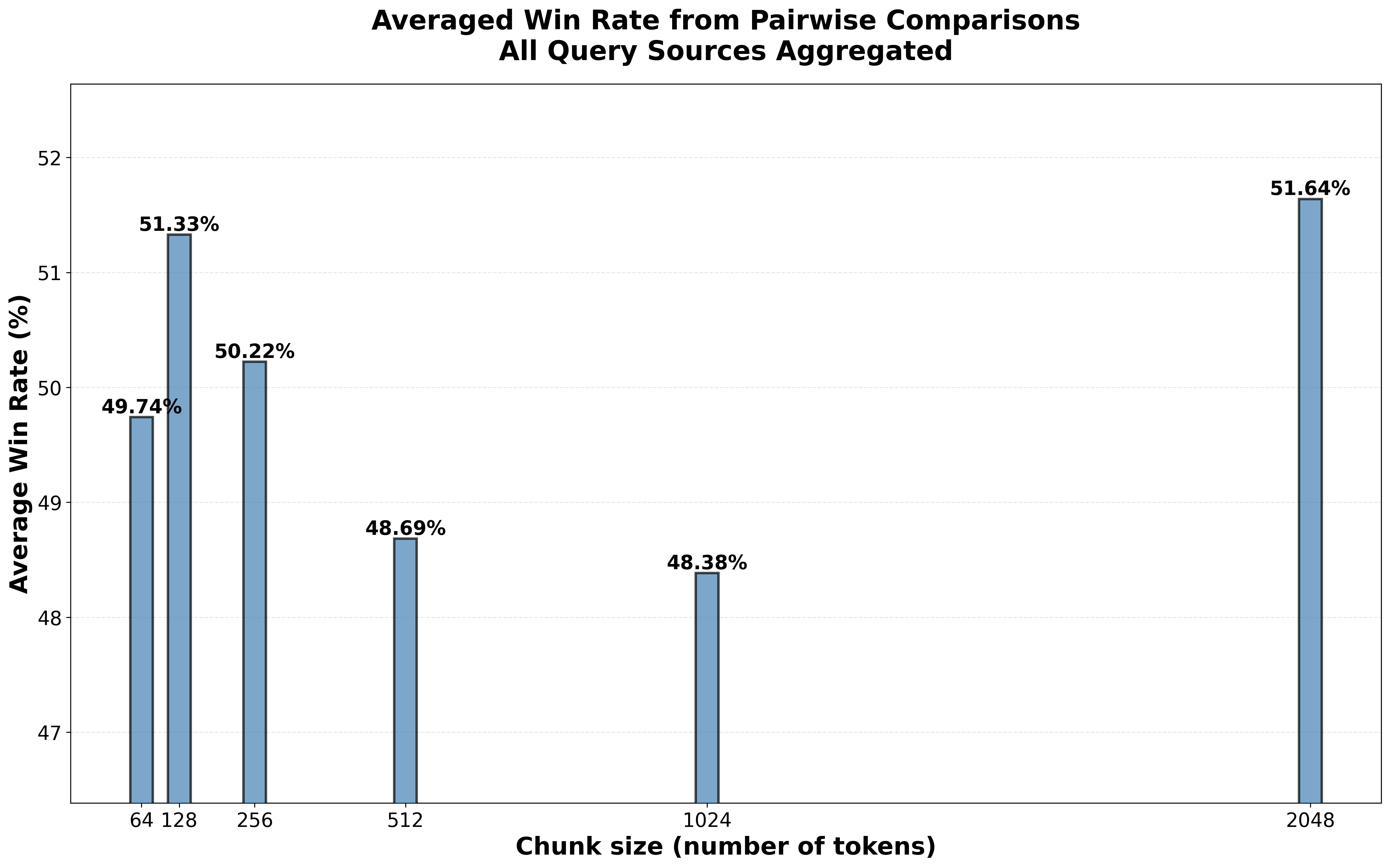}
\caption{Averaged pairwise chunk comparison using rank-aligned win rates aggregated across all query sources. The experiment is run on the PMC Open Access dataset using the \texttt{IndexFLAT} FAISS Index.}
\label{fig:granularities_tournament}
\end{figure}

Figure \ref{fig:granularities_tournament} represents the averaged pairwise chunk comparison results across all query sources.

\section{Example Queries Per-Query-Type}\label{apd:queries}
Table \ref{tab:query_examples} reports representative examples for each query source used in the evaluation of the biomedical retrieval pipeline.

\begin{table*}[t]
\centering
\small 
\floatconts
  {tab:query_examples}
  {\caption{Representative examples for each query source used in the evaluation.}}
  {\begin{tabular}{lp{10cm}} 
  \toprule
  \bfseries Query Source & \bfseries Example Query \\
  \midrule
  MedMCQA & Which vitamin is supplied from only animal source: Vitamin C; Vitamin B7; Vitamin B12; Vitamin D \\
  \addlinespace
  MedQA & A 20-year-old woman presents with menorrhagia for the past several years. She says that her menses “have always been heavy”, and she has experienced easy bruising for as long as she can remember. Family history is significant for her mother, who had similar problems with bruising easily. The patient's vital signs include: heart rate 98/min, respiratory rate 14/min, temperature 36.1°C (96.9°F), and blood pressure 110/87 mm Hg. Physical examination is unremarkable. Laboratory tests show the following: platelet count 200,000/mm3, PT 12 seconds, and PTT 43 seconds. What is the most likely cause of this patient’s symptoms? \\
  \addlinespace
  ChatDoctor & I woke up this morning feeling the whole room is spinning when i was sitting down. I went to the bathroom walking unsteadily, as i tried to focus i feel nauseous. I try to vomit but it wont come out.. After taking panadol and sleep for few hours, i still feel the same.. By the way, if i lay down or sit down, my head do not spin, only when i want to move around then i feel the whole world is spinning.. And it is normal stomach discomfort at the same time? Earlier after i relieved myself, the spinning lessen so i am not sure whether its connected or coincidences.. Thank you doc! \\
  \addlinespace
  Health StackExchange & Is eating spicy hot (pungent) food (hot chilli peppers etc.) healthy or harmful? \\
  \addlinespace
  PubMed Titles & Physiological differences among isolates of Phytophthora cinnamomi. \\
  \addlinespace
  PMC Question Sentences & There is controversy regarding the localization of BRCA1 and BRCA2 proteins to either nucleus or cytoplasm and whether the expression is present in premeiotic germ cells or can still be expressed in mitotic spermatogonia. \\
  \addlinespace
  Wikipedia Medical Terms & Phytobezoar \\
  \addlinespace
  MeSH Terms & Anti-Bacterial Agents \\
  \addlinespace
  LLM-generated (Q/A) & Can I take ibuprofen with high blood pressure medication? \\
  \addlinespace
  LLM-generated (Non-QA) & Persistent cough, weight loss, night sweats \\
  \bottomrule
  \end{tabular}}
\end{table*}

\section{Human Validation of LLM-Based Judgments}
\label{apd:human_validation}
To assess the reliability of LLM-based preference judgments, we conducted a human validation study on a randomly sampled subset of retrieval comparisons.
We uniformly sampled 100 passage-level comparisons from the full evaluation set, spanning multiple query types and retrieval configurations.

Each comparison consisted of a biomedical query and two retrieved passages at the same rank position, produced by different retrieval settings.
Two human annotators with biomedical domain familiarity independently judged which passage was more helpful for answering the query, or indicated a tie when neither passage was clearly preferable.
Annotators were blinded to the identity of the retrieval systems.

We measure agreement between the LLM-based judge and human annotations by computing the fraction of comparisons for which the LLM judgment matched the majority human preference.
Across the 100 sampled comparisons, the LLM judge agreed with the human annotators in 87\% of cases, indicating strong alignment between automated and human preference judgments.
\end{document}